\documentclass[a4paper,11pt]{article}
\usepackage{jcappub}

\newcommand{\Mpl}{M_{\textrm{Pl}}}
\renewcommand{\(}{\left(}
\renewcommand{\)}{\right)}

\def\e{\mathrm{e}}

\def\doi{http://doi.org}
\def\arxiv{http://arxiv.org/abs}

\title{\boldmath Tachyon field non-minimally coupled to massive neutrino matter}

\author[a]{Safia Ahmad}
\author[b]{Nurgissa Myrzakulov}
\author[b]{R. Myrzakulov}

\affiliation[a]{Centre for Theoretical
Physics, Jamia Millia Islamia,\\ New Delhi-110025, India}
\affiliation[b]{Eurasian  International Center for Theoretical Physics, Eurasian National University,\\ Astana 010008, Kazakhstan}

\emailAdd{safia@ctp-jamia.res.in}
\emailAdd{nmyrzakulov@gmail.com}
\emailAdd{rmyrzakulov@gmail.com}

\abstract{In this paper, we consider rolling tachyon, with steep run-away type of potentials non-minimally coupled to massive neutrino matter. The coupling dynamically builds up at late times as neutrino matter turns non-relativistic. In case of scaling and string inspired
potentials, we have shown that  non-minimal coupling leads to
minimum in the field potential. Given a suitable choice of model parameters, it is shown to give rise to late-time acceleration with the desired equation of state.}

\keywords{dark energy theory, string theory and cosmology}

\begin{document} 
\maketitle
\flushbottom

\section{Introduction}
Since the discovery of late-time cosmic acceleration in 1998 \cite{SN,SN2,SN3}, model
building was undertaken in cosmology to capture this remarkable
phenomenon. To this effect, a variety of scalar field models
including quintessence \cite{Ratra:1987rm,Peebles:1987ek,Copeland:2006wr,Sahni:1999gb,
Frieman:2008sn,Padmanabhan:2002ji,Padmanabhan:2006ag,
Sahni:2006pa,Peebles:2002gy,Perivolaropoulos:2006ce,
Sami:2009dk,Sami:2009jx,Sami:2013ssa}, phantom
field \cite{Caldwell:1999ew,Caldwell:2003vq,Carroll:2003st,Singh:2003vx,Hao},
rolling tachyon \cite{Sen:2002nu,Sen:2002in,Sen:2003mv,Mazumdar:2001mm} and others
were investigated in the literature. This description, however, has
limited predictive power. Indeed, for {\it a priori} given cosmic history,
one can always construct a field potential which would give rise to
that history. However, cosmological dynamics of scalar field may
still be of interest if the field has generic features such as
tracking behavior \cite{Ratra:1987rm,Peebles:1987ek} or the field is inspired by a fundamental theory
of high energy physics. For instance, rolling tachyon, inspired by
string theory, was taken up with great interest in cosmology few
years back. The string inspired potential for rolling tachyon is
typically run-away type. It behaves like decaying exponential away
from the maximum such that dark matter is a late-time attractor of
the dynamics \cite{Copeland:2004hq,Gibbons:2002md,Padmanabhan:2002cp,Bagla:2002yn,Padmanabhan:2002sh,
Aguirregabiria:2004xd,Frolov:2002rr,Rangdee:2012rw,Hao:2002qw,Gorini:2003wa,Chimento:2003ta,
Causse:2003hp,Abramo:2004ji,Srivastava:2004wy,Yang:2012ht,Shchigolev:2012jx,Elizalde:2004gp,
Landim:2015poa,Landim:2015uda}. Following work of Sen \cite{Sen:2002nu,Sen:2002in}, efforts were made to obtain
inflation \cite{Fairbairn:2002yp,Choudhury:2002xu,Sami:2002fs,Sami:2002zy,Feinstein:2002aj,Shiu:2002qe,
Kofman:2002rh,Hwang:2002fp,Piao:2002vf,Piao:2002nh,Cline:2002it,Mukohyama:2002vq,Felder:2002sv,
  Bento:2002np,Li:2002et,Kim:2002zr,Matsuda:2003ej,Das:2003xw,Guo:2003zf,Gibbons:2003gb,
  Majumdar:2003kd,Nojiri:2003ag,Elizalde:2003ku,Garousi:2004uf,Sami:2003qx,Garousi:2004hy,
  Felder:2004xu,Calcagni:2004ug,Raeymaekers:2004cu,Calcagni:2004as,Barnaby:2004kz,Panda:2005sg,
  Panigrahi:2004qr,Ghodsi:2004wn,Shiu:2002xp,Shiu:2002cb,Paul:2003jx,Steer:2003yu,
  Nozari:2013mba,Mukohyama:2002cn} using rolling tachyon. Though there is region of slow-roll
near the top of the potential, one could not collect enough number
of e-folds; there is no adjustable parameter in the potential (see also Ref. \cite{Barbosa-Cendejas:2015rba} on the related theme).

Following the initial development, cosmologists investigated
phenomenologically a number of potentials in the framework of
rolling tachyon. For instance, it was shown that $V(\phi)\sim
\phi^{-2}$, an analog of exponential potential for standard
quintessence, gives rise to scaling solution as an attractor of the
dynamics \cite{Ratra:1987rm,Peebles:1987ek}. The equation of state parameter for rolling tachyon is
given by, $\omega_\phi=(\dot{\phi}^2-1)$ which remains constant
during scaling regime such that
 $\omega_\phi=\omega_b$ \cite{Copeland:2004hq,Aguirregabiria:2004xd,Tsujikawa:2004dp,tapan05}. Thus, it requires
background fluid with negative equation of state $\omega_b$ thereby
the scaling solution can not track the standard
background(matter/radiation) in this case. Be it the string inspired
case or the scaling potential, they can not account for late-time
acceleration. It was demonstrated in Ref.~\cite{tapan05} that
coupling of tachyon with matter gives rise to scaling solution which
is accelerating and mimics standard matter.

During slow-roll, rolling tachyon dynamics reduces to canonical
description. Coupling to matter then induces minimum in the
potential. For large value of the coupling, minimum of the potential
is more pronounced and the field can easily be trapped there
mimicking cosmological constant like behavior. Unfortunately, in
this case, the attractor is reached once matter phase is established
which is not a desirable feature; the matter phase should be left
intact.

The above mechanism based upon non-minimal coupling may be salvaged by
invoking non-minimal coupling with massive neutrino matter \cite{Wetterich:2013aca,Wetterich:2013jsa,Wetterich:2013wza,Wetterich:2014eaa,
Wetterich:2014bma,Fardon:2003eh,Bi:2003yr,
Hung:2003jb,Peccei:2004sz,Bi:2004ns,Brookfield:2005td,Brookfield:2005bz,
Amendola:2007yx,Bjaelde:2007ki,Afshordi:2005ym,Wetterich:2007kr,Mota:2008nj,
Pettorino:2010bv,LaVacca:2012ir,Collodel:2012bp,Motohashi:2012wc,Chudaykin:2014oia,
Hossain:2014xha,Hossain:2014zma} leaving
the standard matter minimally coupled. We should admit that this is
purely a phenomenological setting which looks attractive for the
following reason. Since neutrinos become non-relativistic at late
times, the coupling builds up dynamically at late stages. The late-time acceleration gets associated with
the said physical phenomenon.

In this paper, we shall study rolling tachyon, with scaling and
string inspired potentials, assuming non-minimal coupling to massive
neutrino matter and examine the possibility of late time
acceleration in these cases.

\section{Rolling tachyon and late time dynamics}
We shall consider a situation where a tachyon field $\phi$, with an
equation of state $\omega_{\phi} \equiv p_{\phi}/\rho_{\phi}$, is
coupled non-minimally to massive neutrino matter with an equation of
state $\omega_{\nu} \equiv p_{\nu}/\rho_{\nu}$. The continuity
equations for $\rho_{\phi}$ and $\rho_{\nu}$ in the spatially flat
Friedmann-Robertson-Walker (FRW) background are
\cite{piazzashinji04}
\begin{eqnarray}
\label{rhoeqn}
\dot{\rho_\phi}+3H(\rho_\phi+p_\phi)=-Q\sqrt{V}(\rho_\nu-3p_\nu)\frac{\dot{\phi}}{\Mpl},\\
\dot{\rho_\nu}+3H(\rho_\nu+p_\nu)=Q\sqrt{V}(\rho_\nu-3p_\nu)\frac{\dot{\phi}}{\Mpl},
\end{eqnarray}
where Q is the coupling between the tachyon field and the neutrino matter \cite{Amendola:1999er}, $\Mpl$ is the reduced
Planck mass and
dots are the derivatives with respect to the time $t$.\\
The Lagrangian of a tachyon field is given by
\begin{equation}
\label{tpd}
p_{\phi}=-V(\phi)\sqrt{1-\dot{\phi}^2}
\end{equation}
where we have chosen an inverse square potential, $V(\phi)=4
\Mpl^2/(\lambda^2\phi^2)$. Here $\lambda$, which is related to the
slope of the potential, can be determined for scaling solution,
$i.e.$ when $\rho_{\phi}/\rho_{\nu}=$ const, by following the same
procedure as in \cite{piazzashinji04,Tsujikawa:2004dp} and taking
into account the change in the total fractional density at late
times for the system being considered here, which is
$\Omega_m+\Omega_{\phi}+\Omega_{\nu}=1$,
\begin{eqnarray}
\lambda \equiv -\Mpl\frac{V'}{V^{3/2}} = Q\frac{(1-3\omega_{\nu})(\Omega_{\nu}(1+\omega_{\nu})+\Omega_{\phi}(1+\omega_{\phi}))}{\Omega_{\phi}(\omega_{\nu}-\omega_{\phi})}
\end{eqnarray}
where, $\Omega_{\phi}$ is the tachyon field density parameter.\\
The energy density for the field $\phi$ corresponding to the
Lagrangian (\ref{tpd}) is
\begin{equation}
\label{ted}
\rho_{\phi}=\frac{V(\phi)}{\sqrt{1-\dot{\phi}^2}}
\end{equation}
Equation of state parameter for tachyon field is therefore,
\begin{equation}
\label{wphi}
\omega_{\phi}\equiv \frac{p_{\phi}}{\rho_\phi}=\dot{\phi}^2-1
\end{equation}
In the spatially flat Friedmann-Robertson-Walker background, we have
the following Friedmann equations,
\begin{eqnarray}
3H^2 \Mpl^2 &=& \rho_m+\rho_r+\rho_\phi+\rho_\nu\, ,\\
(2\dot{H}+3H^2)\Mpl^2 &=& V(\phi)\sqrt{1-\dot{\phi}^2}-\omega_\nu \rho_\nu
\end{eqnarray}
Here we have used the standard continuity equations of radiation and non-relativistic matter (excluding massive neutrino matter which couples directly to the tachyon field),
\vspace{-.01in}
\begin{eqnarray}
\dot{\rho_m}+3H\rho_m =0\\
\dot{\rho_r}+4H\rho_r =0
\end{eqnarray}
The equation of motion for the tachyon field $\phi$ can then be obtained by using the eqns.~(\ref{rhoeqn}),(\ref{ted}) and (\ref{wphi}),
\begin{equation}
\frac{\ddot{\phi}}{1-\dot{\phi}^2}+3H\dot{\phi}+\frac{V^{'}}{V}=-\frac{\sqrt{1-\dot{\phi}^2}}{\sqrt{V} \Mpl}Q(\rho_\nu-3p_\nu)
\end{equation}
with the equation-of-state parameter for the
neutrino matter defined in such a way that neutrinos are relativistic for most of
the history of the universe and become non-relativistic in the recent times \cite{Hossain:2014xha},
\begin{equation}
\omega_\nu=\frac{1}{6}\left[1+\tanh\left(\frac{\ln(1+z)-z_{\rm eq}}{z_{\rm dur}}\right)\right]
\end{equation}
where the epoch around which the transition of neutrino from relativistic to non-relativistic occurs, is determined by $z_{\rm eq}$ and $z_{\rm dur}$ determines how smoothly this
transition takes place.\\
We define the following dimensionless energy density
parameters for matter, radiation, neutrino and tachyon field, respectively,
\begin{eqnarray}
\Omega_m=\frac{\rho_m}{3H^2\Mpl^2}\, \\
\Omega_r=\frac{\rho_r}{3H^2\Mpl^2}\, \\
\Omega_{\nu}=\frac{\rho_{\nu}}{3H^2\Mpl^2}\, \\
\Omega_{\phi}=\frac{\rho_{\phi}}{3H^2\Mpl^2}
\end{eqnarray}
\begin{figure}
\includegraphics{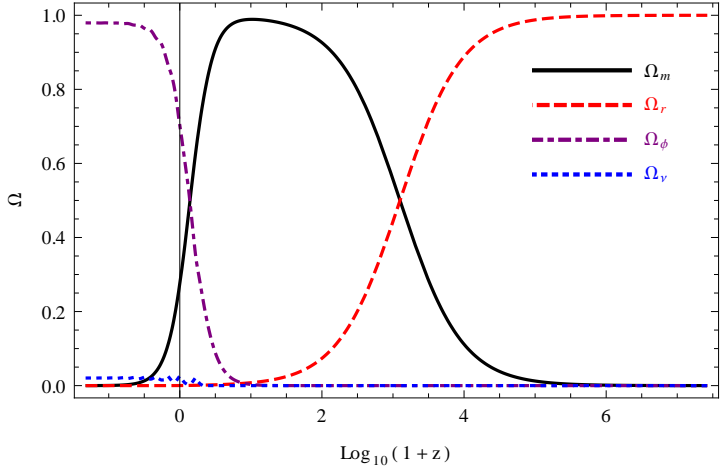}
\caption{\label{fig:omegas} Figure shows the evolution of energy density parameters for radiation(red dashed line), matter(black solid line), tachyon field(purple dotted line) field and neutrino matter(blue dotdashed line), respectively, with respect to $\rm Log_{10}(1+z)$. Here, we have considered $z_{\rm {dur}}=3.0$.}
\end{figure}
\begin{figure}
\includegraphics{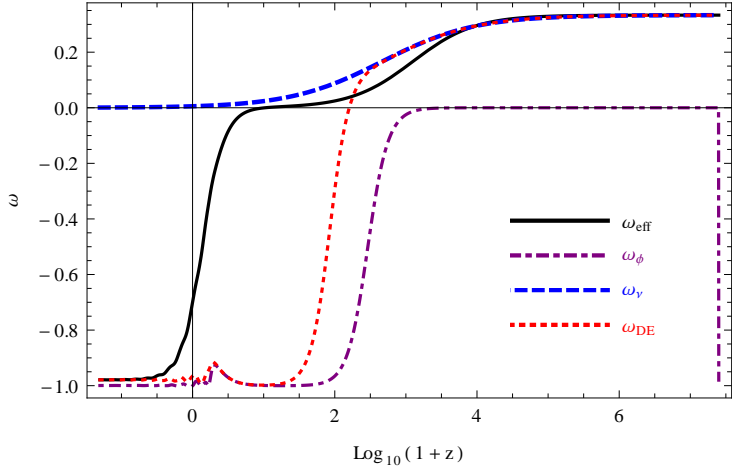}
\caption{\label{fig:eos} Figure shows the evolution of equation-of-state parameters, $\omega_{\rm eff}$ (black solid line), $\omega_{\phi}$ (purple dotdashed line), $\omega_{\nu}$ (blue dashed line) and $\omega_{\rm DE}$ (red dotted line), respectively, with respect to $\rm Log_{10}(1+z)$. We took $z_{\rm {dur}}=3.0$ to plot this figure.}
\end{figure}
Figure~\ref{fig:omegas} shows the evolution of density parameters for matter, radiation, neutrino and tachyon field. We have started the analysis of the evolution of Universe from the radiation dominated era followed by the matter dominated epoch. The Universe has recently entered the dark energy dominated epoch which, in this figure, is characterized by the non-minimal coupling of tachyon field with neutrino matter. In figure~\ref{fig:eos}, we have shown the evolution of various equation-of-state parameters as the universe evolves. During the radiation dominated era, since the neutrinos are relativistic, the equation of state parameter for radiation as well as the neutrino matter is 1/3. Neutrinos remain relativistic even after the radiation cease to dominate and have become non-relativistic only in the recent past. The equation-of-state parameters, $\omega_{\rm eff}$, $\omega_{\phi}$ and $\omega_{\rm DE}$, approaches -1 at the present time since the universe is going through an accelerated expansion.\\
We define the following dimensionless quantities,
\begin{eqnarray*}
x=\frac{\dot{\phi}}{\sqrt{2}}\, ,~~y=\frac{\sqrt{V}}{\sqrt{3}H\Mpl}\, ,~~\Omega_m=\frac{\rho_m}{3H^2\Mpl^2}\, ,~~\Omega_r=\frac{\rho_r}{3H^2\Mpl^2}\, ,\\
\lambda=-\Mpl\frac{V'}{V^{3/2}}\, ,~~\omega_\nu=\frac{1}{6}\left[1+\tanh\left(\frac{\ln(1+z)-z_{\rm eq}}{z_{\rm dur}}\right)\right]
\end{eqnarray*}
so that the cosmological equations in the autonomous form are written as,
\begin{eqnarray}
\label{xn}
\frac{{\rm d}x}{{\rm d}N} &=& (1-2x^2) \left(-3x+\sqrt{\frac{3}{2}}\lambda y
+\sqrt{\frac{3}{2}}\frac{\sqrt{1-2 x^2}}{y}Q(3\omega_\nu-1)\Omega_\nu \right) ,\\
\label{yn}
\frac{{\rm d}y}{{\rm d}N} &=& \frac{y}{2} \left(3-\sqrt{6}\lambda xy-3 y^2\sqrt{1-2x^2}+3\omega_\nu\Omega_\nu+\Omega_r \right) ,\\
\label{omn}
\frac{{\rm d}\Omega_m}{{\rm d}N} &=& \Omega_m \left(-3 y^2\sqrt{1-2x^2}+3\omega_\nu \Omega_\nu+\Omega_r \right) ,\\
\label{orn}
\frac{{\rm d}\Omega_r}{{\rm d}N} &=& \Omega_m \left(-1-3 y^2\sqrt{1-2x^2}+3\omega_\nu \Omega_\nu+\Omega_r \right) ,\\
\label{wnun}
\frac{{\rm d}\omega_\nu}{{\rm d}N} &=& \frac{2\omega_\nu}{z_{\rm dur}}(3 \omega_{\nu }-1)
\end{eqnarray}
where, $N \equiv \ln a$, is the number of e-foldings.\\
The equation of state parameters in terms of the dimensionless variables can now be written as
\begin{eqnarray*}
\omega_\phi &=& 2x^2-1\, ,\\
\omega_{\rm eff} &=& -y^2\sqrt{1-2 x^2}+\omega_\nu\Omega_\nu+\frac{\Omega_r}{3}
\end{eqnarray*}
Energy density parameter for the tachyon field in terms of the dimensionless variables can be written as
\begin{equation}
\Omega_{\phi}=\frac{y^2}{\sqrt{1-2x^2}},
\end{equation}
and since
\begin{equation}
\Omega_m+\Omega_r+\Omega_{\phi}+\Omega_{\nu}=1,
\end{equation}
we can write the energy density parameter for neutrino matter as
\begin{equation}
\Omega_{\nu}=1-\Omega_m+\Omega_r+\Omega_{\phi}
\end{equation}
Dark energy contribution comes from the tachyon field as well as the neutrino matter
\begin{equation}
\Omega_{DE}=\Omega_{\phi}+\Omega_{\nu}
\end{equation}
Fixed points can be obtained by equating eq.~(\ref{xn})-(\ref{wnun})
to zero. The fixed points for the system in which a tachyon field is non-minimally
coupled to matter are derived in \cite{tapan05} and will be the same if we consider the case of tachyon field non-minimally coupled with neutrino matter.  We are
interested in the one which corresponds to the scaling solution,
which as given in \cite{tapan05} is,
\begin{equation}
x=\bar{x},~~y=\frac{3}{\sqrt{6}(\lambda+Q)\bar{x}}
\end{equation}
where, $\bar{x}$ satisfies the following relation
\begin{equation}
\frac{\sqrt{1-2\bar{x}^2}}{\bar{x}^2}=\frac{2Q(\lambda+Q)}{3}
\end{equation}
The equation of state parameter for the tachyon field corresponding to the above fixed point is,
\begin{eqnarray}
\omega_{\phi} = 2\bar{x}^2-1 = \frac{9}{2Q^2(Q+\lambda)^2}\left[-1+\sqrt{1+\frac{4}{9}Q^2(Q+\lambda)^2}\right]-1
\end{eqnarray}
Clearly, as $Q \rightarrow 0$, $x \rightarrow 1/\sqrt{2}$, $y \rightarrow \sqrt{3}/(\lambda+Q)$, $\omega_{\phi}  \rightarrow 0$ while in the limit $Q  \rightarrow \infty$, $x \rightarrow 0$, $y \rightarrow 1$, $\omega_{\phi} \rightarrow -1$. So, the accelerated expansion is realised for large Q.\\
From eq.~(\ref{wphi}), it can be clearly seen that late time acceleration
can only happen if $\dot{\phi}$ is very small. So in the
limit when $\dot{\phi}\ll1$, tachyon lagrangian (\ref{tpd}) can be approximated at late times as,
\begin{equation}
\label{lagbinexp}
p_{\phi}\approx -V(\phi)+V(\phi)\frac{\dot{\phi}^2}{2}
\end{equation}
which can be written in the canonical form by redefining the field as
\begin{equation}
\label{phisig}
V(\phi)=\left(\frac{\partial \sigma}{\partial \phi} \right)^2,
\end{equation}
provided that the potential is very shallow at late times.
Relation between $\phi$ and $\sigma$ can be obtained by integrating (\ref{phisig})
\begin{equation}
\frac{\lambda\sigma}{2\Mpl}=\ln \phi -\ln C
\end{equation}
so that $V(\sigma)$ reads
\begin{equation}
V(\sigma)=\frac{4\Mpl^2}{\lambda^2C^2}e^{-\lambda \sigma/\Mpl}
\end{equation}
where C is the integration constant.\\
So, if at present time $\sigma=\sigma_0$ and $V(\sigma)=V(\sigma_0)$, then
\begin{equation}
C^2=\frac{4\Mpl^2}{\lambda^2V(\sigma_0)}e^{-\lambda \sigma_0/\Mpl}
\end{equation}
so that
\begin{equation}
V(\sigma)=V(\sigma_0)e^{-\lambda(\sigma-\sigma_0)/\Mpl}
\end{equation}
Eq.~(\ref{lagbinexp}) becomes
\begin{equation}
p_{\sigma}=\frac{\dot{\sigma}^2}{2}-V(\sigma),
\end{equation}
with equation of state parameter taking the form
\begin{equation}
\omega_{\sigma}=\frac{\dot{\sigma}^2}{V}-1
\end{equation}
and equation of motion in terms of the new field $\sigma$ is
\begin{equation}
\ddot{\sigma}+3H\dot{\sigma}=-\frac{\partial V}{\partial \sigma}-\frac{Q}{\Mpl}\rho_{\nu}
\end{equation}
considering the fact that neutrinos become non-relativistic at late times.\\
Now, if we define, $\tilde{\rho_{\nu}}=\rho_{\nu}e^{-Q (\sigma-\sigma_0)/\Mpl}$, then the effective potential can be written as
\begin{equation}
V_{\rm eff}(\sigma)=V(\sigma)+\tilde{\rho_{\nu}}e^{Q (\sigma-\sigma_0)/\Mpl}
\end{equation}
which has a minimum at
\begin{equation}
\sigma_{\rm min}=\sigma_0+\frac{\Mpl}{Q+\lambda}\ln\left[\frac{V(\sigma_0)\lambda}{Q \tilde{\rho_{\nu}}}\right]
\end{equation}
which gives us the effective potential, $V_{\rm eff}$ at the minimum
\begin{equation}
V_{\rm eff}(\sigma_{\rm min})=V(\sigma_{\rm min})\left[1+\frac{\lambda}{Q} \right]
\end{equation}
where,
\begin{equation}
V(\sigma_{\rm min})=\frac{Q}{\lambda}\rho_{\nu}(\sigma_{\rm min}),
\end{equation}
is the potential at minimum. Now, effective potential can be rewritten as
\begin{equation}
\label{eq:Veff_min}
V_{\rm eff}(\sigma_{\rm min})=\Omega_{\nu}(\sigma_{\rm min})\left[1+\frac{Q}{\lambda} \right]3H_0^2\Mpl^2
\end{equation}
As given by the Planck 2015 results \cite{Ade:2015xua}, the bound on neutrino matter is $\Omega_{\nu} \lesssim 0.005$ and the fact that field must settle down at the minimum of the effective potential in order to get the late time cosmic acceleration, $Q\gg\lambda$ for the effective potential to be of the order of $3 H_0^2\Mpl^2$ at the minimum.\\
Furthermore, we have determined the cosmological deceleration-acceleration  transition  redshift \cite{Farooq:2013eea,Farooq:2013hq,Capozziello:2014zda,Farooq:2013dra,Farooq:2012ev}. The deceleration parameter, for the model under consideration, is defined as
\begin{eqnarray}
q \equiv -\frac{\ddot{a}}{a H^2} = \frac{1}{2} \left[ \Omega_m + \Omega_{\phi} (1+3\omega_{\phi})+ \Omega_{\nu} (1+3\omega_{\nu}) + 2\Omega_r \right]
\end{eqnarray}

\begin{figure}
\includegraphics{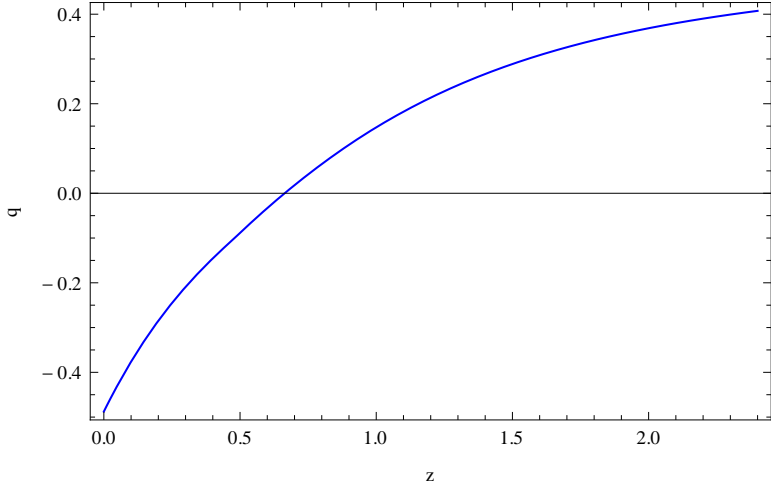}
\caption{\label{fig:Deceleration} Figure shows the evolution of deceleration parameter, $q$ with respect to redshift $z$.}
\end{figure}

The redshift at which the deceleration parameter, $q$, changes sign from positive to negative corresponds to the onset of late-time acceleration. Figure~\ref{fig:Deceleration} shows the evolution of the deceleration parameter $q$ with respect to the redshift $z$. The redshift around which the transition from the decelerating expansion to the accelerating expansion occurs is found to be $z_{\rm {da}}=0.663$.

\subsection{Observational Constraints}

\begin{figure}
\includegraphics{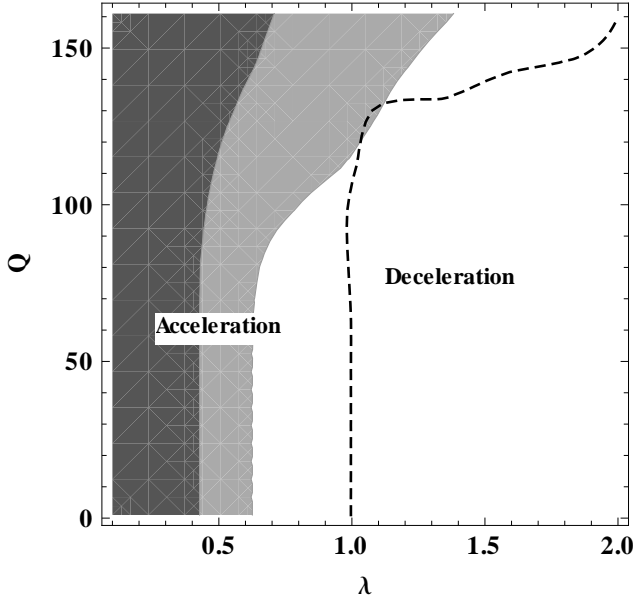}
\caption{\label{fig:contour} Figure shows the 1$\sigma$ (grey) and 2$\sigma$ (light grey) contours for the model parameters $Q$ and $\lambda$ using SN1a and BAO data. The dashed line corresponds to the onset of late-time cosmic acceleration. The $q=0$ dashed line clearly separates the decelerating and accelerating regimes.}
\end{figure}

\begin{figure}
\includegraphics{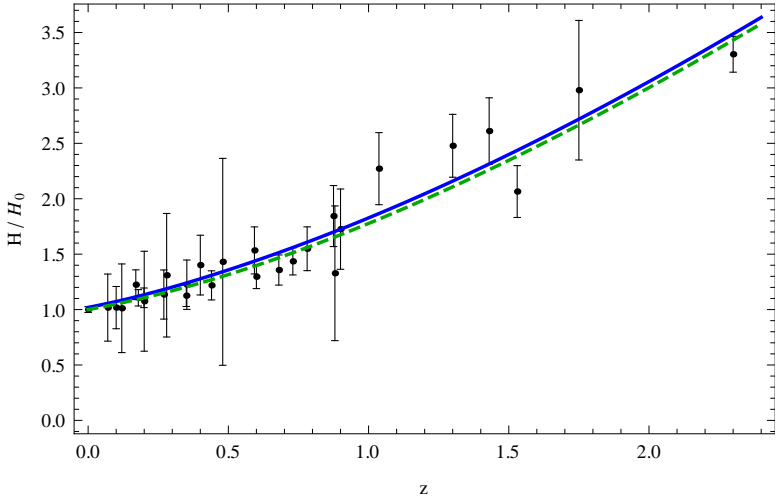}
\caption{\label{fig:hubble} Figure shows the normalized Hubble parameter for the model being studied in this paper(blue solid line) and $\Lambda$CDM(green dashed line) with respect to the redshift, $z$. We have also plotted the normalized Hubble data with 1$\sigma$ error bars calculated from the  compilation of H(z) data  points \cite{Simon:2004tf,Stern:2009ep,Zhang:2012mp,Moresco:2012by,Chuang:2012qt,Blake:2012pj,
Busca:2012bu}.}
\end{figure}

In figure~\ref{fig:contour}, we have constrained our model parameters $(Q,\lambda)$ using Supernovae and BAO data. We have used the compilation of 580 data points of Union2.1 dataset \cite{Suzuki:2011hu} for SN1a observation and for BAO observation, we have used the data from \cite{Giostri:2012ek}.
The constraints on the model parameter $Q$ and $\lambda$ are obtained by marginalising $\chi^2$ over the value of $\Omega_{\nu 0}$. As can be seen from the figure, the values of $\lambda > 0.72$ are ruled out at 1$\sigma$  level while from 2$\sigma$ confidence contour, the upper bound on $\lambda$ is $1.4$ for large vales of $Q$ so that the ratio $Q/ \lambda \gg 1$ which is required for the minimum of the effective potential to be of the order of $3 H_0^2\Mpl^2$ at the present epoch. The dashed line in the plot corresponds to the redshift where $q=0$.\\
In figure~\ref{fig:hubble}, we have shown the evolution of normalized Hubble parameter for our model and compared it with that of the $\Lambda$CDM model. We have also plotted data points for normalized Hubble parameter, $H/H_0$ with 1$\sigma$ error bars which have been calculated from the compilation of 29 points of H(z) data \cite{Farooq:2013eea,Farooq:2013hq} using the present value of Hubble parameter, $H_0=(67.8 \pm 0.9)$ km s$^{-1}$ Mpc$^{-1}$ from Planck 2015 results \cite{Ade:2015xua}. Error in $H/H_0$ can be calculated \cite{Bamba:2013aca} as
\begin{equation}
\sigma_{(H/H_0)} = \( \frac{\sigma_H}{H} + \frac{\sigma_{H_0}}{H_0} \) \frac{H}{H_0}
\end{equation}
We thereby concludes that our model satisfies the observational constraints in the comfortable region of the parameter space $(Q,\lambda)$.

\section{Cosmology with string inspired Potential}

\begin{figure}
\includegraphics{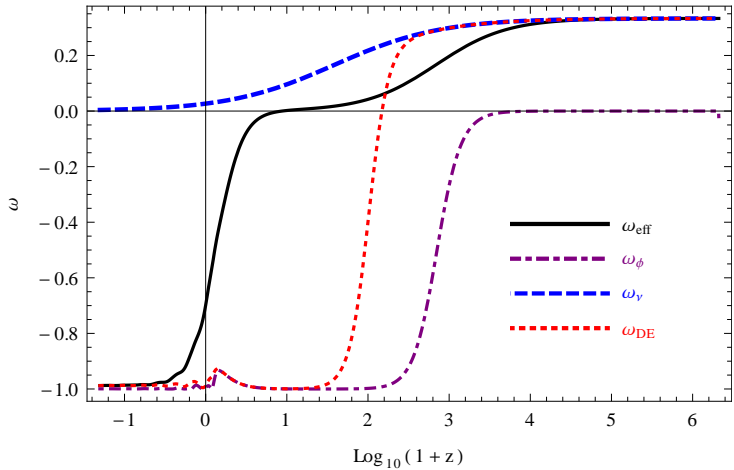}
\caption{\label{fig:weff} Figure shows the evolution of equation-of-state parameter with respect to $\rm Log_{10}(1+z)$, for the case of string inspired exponential potential.}
\end{figure}
As mentioned in the introduction, the string inspired potential
for rolling tachyon typically behaves as decaying exponential away
from its maximum. In this case, the dark matter like solution with
$\omega_\phi=0$ is a late-time attractor of the system \cite{Sami:2002fs}. In what follows,
 we
now consider the case of exponential potential of tachyon field of
the form
\begin{equation}
V(\phi)=V_0 \e^{-\alpha \phi}
\end{equation}
where $\alpha$ is a constant with mass dimension 1 and $\lambda$ is the
dimensionless parameter which is associated with the
slope of potential. In case of the exponential potential, we
have
\begin{eqnarray}
\lambda \equiv -\Mpl \frac{V'}{V^{3/2}} = \frac{\alpha\Mpl}{\sqrt{V}}
\end{eqnarray}
which is not constant and hence does not give the scaling solution.
Rolling tachyon with string inspired potential as such can not give
rise to late-time acceleration. One might bring in a class of
phenomenological potentials to do the job. In our opinion, it could
still be of interest if the dynamics has generic features such as
scaling behavior we considered before. As for the string inspired
case, we could hope to get the desired late-time behavior by bringing
in again the massive neutrino matter.
 Considering then the interaction between tachyon
field and massive neutrino matter, we study the late-time dynamics for
exponential potential by following the same procedure as above. So,
in addition to the cosmological equations (\ref{xn})-(\ref{wnun}), we
also have the evolution equation of $\lambda$,
\begin{equation}
\frac{{\rm d}\lambda}{{\rm d}N} = \sqrt{\frac{3}{2}}\lambda^2 xy
\end{equation}
Solving these equations numerically, the figure~\ref{fig:weff} shows
that even though we get non-scaling solutions, the effective
equation of state parameter approaches -1 at late times because of
the coupling between tachyon field and neutrino matter.

\section{Conclusion}
In this paper, we have investigated a system in which the tachyon
field is non-minimally coupled to massive neutrino matter. In such a
system, the equation of state parameter of neutrino matter is chosen
phenomenologically such  that neutrinos are relativistic for most of
the history of universe and become non-relativistic only in the
recent past. We have considered potentials that correspond to
scaling as well as the non-scaling solutions. First, we have
investigated  the inverse square potential, $\phi^{-2}$, for which
the slope, $\lambda=-\Mpl V'/V^{3/2}$ is a constant thereby the
tachyon field exhibits a scaling behavior. However, in this case, the field can mimic
a hypothetical background fluid with negative equation of state. We have shown that
situation can be remedied by invoking non-minimal coupling to massive neutrino matter.\\
In presence of
coupling, the scaling solution is an attractor of the system which
mimics the desired equation of state for a very large value of
coupling constant, $Q$. We numerically checked that
 $\dot{\phi} \ll 1$ in the asymptotic limit. In this limit,  it is shown that
the tachyon field is reduced to the canonical field at late times
through a suitable transformation. Since tachyon is non-minimally
coupled to neutrino matter and coupling builds up dynamically only
at late stages, the effective potential of the field
acquires a minimum at late times. \\
According to  Planck 2015 results \cite{Ade:2015xua},
 the bound on present value of massive neutrino matter is $\Omega_{\nu} \lesssim 0.005$ which implies that coupling should be large for the field  to settle down at the minimum of the potential around the  present epoch(as shown in eq.~(\ref{eq:Veff_min})).
 We have analytically demonstrated that late time cosmic acceleration can be realized if the tachyon field is coupled
 non-minimally to the massive neutrino matter thereby leaving the matter dominated epoch intact. We have also shown the 1$\sigma$ and 2$\sigma$ confidence level in the parameter space $(Q,\lambda)$ using SN1a and BAO data. The normalized Hubble parameter for $\Lambda$CDM and our model have also been plotted and we have compared that with the H(z) data. \\
Secondly, we have also examined the case of string inspired exponential potential that does not give rise to the scaling solution for the model under consideration. Indeed, in this case, the dark matter like solution is a late-time attractor of the system. We have demonstrated that due to the coupling between tachyon and massive neutrino matter, $\omega_{\rm eff}$  approaches -1 at late times as shown in figure~\ref{fig:weff}. We therefore conclude that rolling tachyon with and without scaling potential can successfully give rise to late time acceleration by invoking non-minimal coupling with massive neutrino matter.

\section*{ACKNOWLEDGMENTS}
We are thankful to M. Sami and Md. Wali Hossain for useful discussions. Safia Ahmad acknowledges DST, Govt. of India for financial support through Inspire Fellowship (DST/INSPIRE Fellowship/2012/614).

\end{document}